# Resonance Damping of the THz-frequency Transversal Acoustic Phonon in the Relaxor Ferroelectric KTa$_{1-x}$Nb$_x$O$_3$

J.Toulouse[1] E.Iolin[1,2], B.Hennion[3], D.Petitgrand[3], and R.Erwin[4]

1 Physics Department, Lehigh University, Bethlehem, PA, USA, jt02@lehigh.edu
2 Latvian Academy of Science, Riga, Latvia, eiolin@netzero.net
3 Laboratoire Leon Brillouin, CEA, Saclay, France
4 National Center for Neutron Research, NIST, Gaithersburg, MD, USA

**Abstract.**
The damping ($\Gamma_a$) of the transverse acoustic (TA) phonon in single crystals of the relaxor KTa$_{1-x}$Nb$_x$O$_3$ with x=0.15-0.17 was studied by means of high resolution inelastic cold neutron scattering near the (200) B.Z. point where diffuse scattering is absent, although it is present near (110). In a wide range of temperatures centered on the phase transition, $T$=195K÷108K, the transverse acoustic (TA) phonon width (damping) exhibits a step increase around momentum $q$=0.07, goes through a shallow maximum at $q$=0.09-0.12 and remains high above and up to the highest momentum studied of $q$=0.16.
These experimental results are explained in terms of a resonant interaction between the TA phonon and the collective or correlated reorientation through tunneling of the off-center Nb$^{+5}$ ions. The observed TA damping is successfully reproduced in a simple model that includes an interaction between the TA phonon and a dispersionless localized mode (LM) with frequency $\omega_L$ and damping $\Gamma_L$ ($\Gamma_L < \omega_L$), itself coupled to the transverse optic (TO) mode.
Maximum damping of the TA phonon occurs when its frequency $\omega_a \approx \omega_L$. The values of $\omega_L$ and $\Gamma_L$ are moderately dependent on temperature but the oscillator strength, $M^2$, of the resonant damping exhibits a strong maximum in the range $T$~120 K÷150 K in which neutron diffuse scattering near the (110) B.Z. point is also maximum and the dielectric susceptibility exhibits the relaxor behavior. The maximum value of $M$ appears to be due to the increasing number of polar nanodomains. In support of the proposed model, the observed value of $\omega_L$ is found to be similar to the estimate previously obtained by Girshberg and Yacoby. Alternatively, the TA phonon damping can be successfully fitted in the framework of an empirical Havriliak - Negami (HN) relaxation model that includes a strong resonance-like transient contribution.

## 1. INTRODUCTION

KTN is a member of the family of relaxor ferroelectrics, a now well recognized subgroup of highly polarizable compounds with substitutional disorder and off-center ions displaced from their high symmetry lattice site (see review and recent results [1]). Other well-known and extensively studied systems of the same family are the lead compounds PbMg$_{1/3}$Nb$_{2/3}$O$_3$ (PMN) and PbZn$_{1/3}$Nb$_{2/3}$O$_3$ (PZN). With decreasing temperature, correlations develop between the dipoles introduced by the off-center ions, leading to the formation of polar nanoregions (PNRs) below a temperature known as Burns temperature, $T_d$. At a still lower temperature, $T^* < T_d$, the dielectric susceptibility begins to exhibit the characteristic frequency dispersion displayed by relaxor ferroelectrics. This suggests a distinction between quasi-static polar nanoregions (PNRs), resulting from dynamical polar correlations, and polar nanodomains (PNDs) [2] when these correlations become static or long lived, resulting in local distortions evidenced by elastic diffuse scattering shown below. The existence of quasi-static polar nanoregions (PNRs) in the intermediate temperature range $T^* < T < T_d$ has been recognized for a long time and was initially thought confirmed by neutron diffuse scattering (DS). In the past few years however, it has been found that the DS observed in this intermediate temperature range corresponds to quasi-elastic scattering by low energy phonons [1]. "True" elastic DS only appears at a lower temperature than Burns temperature, $T_{dr} < T_d$, where $T_{dr}$ appears to coincide with the temperature $T^*$ at which the static PNDs appear. In the case of PMN for example, $T_d$=620 K but $T_{dr}$=420±20 K, which is indeed very close to $T^*$. Eventually, long range ferroelectric order develops at a lower Curie temperature, $T_c < T_{dr}$. These recent findings suggest that one should reexamine the original meaning of Burns temperature, $T_d$, and the development of the local polar order. What appears to be well established at least is the presence of precursors in the dynamical pair correlation in the range $T_{dr} < T < T_d$. [1]. New data concerning the genesis of PNRs have recently been reported by Manley et al [3]. These authors have measured inelastic neutron scattering in a single crystal PMN-30%PbTiO$_3$ (PMN-PT) and have found additional intensity between the TA and TO branches, $E_{TA}(q) < E_{RM}(q) < E_{TO}(q)$, exhibiting little dispersion and probably due to localized phonon modes (LM). They have suggested that the interaction between the lattice TO phonon and a low dispersion localized mode is the driving force leading to the appearance of PNRs. However, Gehring et al. [4] have recently shown that the peak observed in their data [3] is spurious and most likely results from a simple double scattering process involving a longitudinal acoustic (LA) mode and strong elastic Bragg scattering at the (2,-4, 0) B.Z. point.

DS in relaxors can exhibit unusual features and its connection to the damping of the TO and TA phonons is still unclear: i) in PMN, Hirota et al. reported strong DS near the (110) BZ point but very weak or almost absent near (200) [5]. They explained their results by proposing that diffuse scattering is not necessarily due to a condensed soft TO mode but can also be due to a *collective shift $\delta$ of all the atoms within a PNR* along the local polar direction, which should certainly have an effect on the transverse acoustic (TA) phonon; ii) also in PMN, Stock et al [6, 7] found DS to be strong near (1,1,0) and accompanied by strong TA damping, but small although still accompanied by moderate TA damping near (2,0,0). These authors therefore concluded to a connection between DS and the damping $\Gamma_a$ of the transverse acoustic (TA) phonon. However, they did not propose a conclusive physical model that would have explained these observations. Recently Phelan *et al.* [1] have shown that the unique electro-mechanical and dielectric features of the relaxor Pb(Mg$_{1/3}$Nb$_{2/3}$)$_{1-x}$Ti$_x$O$_3$ (PMN-xPT), when compared with PbZr$_{1-x}$Ti$_x$O$_3$ (PZT), could be explained by the presence of strong random electric fields (REFs) generated by the heterovalent cations Mg$^{2+}$/Nb$^{5+}$-Ti$^{4+}$ on the B-site in PMNxPT. By contrast, the homovalent Zr$^{4+}$ and Ti$^{4+}$ cations on the B-site in PZT generate only weak random electric fields (REFs). KTN presents an interesting case which puts into question the explanation given above for PMN and PMN-xPT. In KTN, the two alternative cations, Ta and Nb, have the same 5+ valence and the substitution of one by the other does not therefore give rise to random electric fields. As shown below, DS is observed in KTN near the (110) BZ point but is only weak or absent near the (200) BZ point where significant TA damping is nevertheless observed [8]. In the present paper, we examine the cause of this TA damping in a neutron study of KTN.

In KTN, the Ta$^{5+}$ ion is replaced by the homovalent Nb$^{5+}$ ion with almost the same ionic radius, unlike PMN (PbMg$_{1/3}$Nb$_{2/3}$O$_3$) and PZN (PbZn$_{1/3}$Nb$_{2/3}$O$_3$) in which Nb$^{5+}$ replaces the divalent Mg$^{2+}$ and Zn$^{2+}$ ions respectively. Hence, chemical disordering leads to the existence of static random electric fields in PMN and PZN but not in KTN. Moreover, due to the cation valences, Coulomb forces are expected to suppress long-range composition fluctuations in PMN and PZN but to be absent in KTN. KTN is therefore a useful model system for the study of relaxor ferroelectrics with homovalent cations and weak or non-existent REFs. Other such systems are (K$_{1-x}$Li$_x$)TaO$_3$ (KLT), Ba(Zr$_{1-x}$Ti$_x$)O$_3$ (BZT), and Ba(Sn$_{1-x}$Ti$_x$)O$_3$ (BST) etc.. The essential feature of these systems is the presence of isovalent off-center ions and the two types of dynamics in which they participate, one the strictly local motion within the unit cell and the other the correlated/collective motion of off-center ions within PNRs/PNDs coupled to the soft transverse optic (TO) mode. In KTN, the Nb$^{5+}$ cations are displaced from their high symmetry site by 0.145 Å in eight equivalent <111> directions between which they can reorient [9]. These off-center ions create electric dipoles that become correlated at lower temperatures and collectively reorient under the action of an external electric field, giving rise to the characteristic relaxor behavior of the dielectric susceptibility. In addition, these are likely to give rise to localized modes (LM) that can couple to the TO and TA lattice phonons.

In the present paper, we have studied the damping of the TA phonon by means of high resolution inelastic cold neutron scattering near the (200) B.Z. point where DS is very small [8]. We find that the TA damping is small for small wavevectors $q$ but increases rapidly to a maximum value in the vicinity of

$q$~0.1 r.l.u (reciprocal lattice units). We show that this TA damping can be qualitatively explained by either one of two models. In the first model, it is successfully explained in terms of a resonant interaction between the TA phonon mode and a dispersionless localized mode (LM) of frequency $\omega_L$ and damping $\Gamma_L$, $\Gamma_L < \omega_L$, itself strongly coupled to the soft transverse optic (TO) phonon mode (TA-[LM-TO]). The value of the frequency $\omega_L$ is found to be close to the theoretically estimated value for the tunneling splitting of the $Nb^{5+}$ ions in their motion between off-center positions, $\omega_L \sim 30 \div 35K$ [9,10]. Alternatively, the TA damping can be successfully described in terms of an empirical Havriliak-Negami [11] relaxation, revealing a strong resonance-like transient response.

In the following, we present diffuse and inelastic neutron scattering experimental results in section 2, their analysis in section 3, and a general discussion in section 4. The mathematical details of the Havriliak-Negami model are presented in Appendix A. The interaction between tunneling off-center ions among (111) positions and the TO and-TA phonons is formally described in Appendix B. Some earlier results can also be found in references [12] and [13].

## 2. EXPERIMENT

### 2.1. DIFFUSE SCATTERING in THE RELAXOR KTN15.

Diffuse scattering measurements were made on a single crystal $KTa_{0.85}Nb_{0.15}O_3$ (KTN15). The development of a local structural order (PNDs), with lower symmetry than the surrounding lattice, was first evidenced in the KTN15 crystal through elastic diffuse scattering (DS) (within the limit of resolution). DS measurements were carried out on the BT2 spectrometer at the neutron facility at NIST with collimations of 60'-40'-40'-80' and neutron energy of 14.7 meV. Pyrolytic graphite was used to filter out harmonics. The scattering plane was (100)-(011) and elastic scattering was measured around the (110) Bragg peak in transverse <001> scans and upon warming. The results are presented in Fig.1a on two different scales, highlighting the diffuse scattering on a reduced vertical scale and the full scale Bragg intensity as an inset. These elastic scattering results were fitted with two peaks, a Gaussian for the Bragg and a Lorentzian for the diffuse scattering. DS is clearly visible and seen to grow below 160K. The Bragg peak intensity increases rapidly below 140K due to the relief of primary and secondary extinction caused by atomic plane distortions. In Fig.1b, the integrated intensity of the DS is seen to reach a maximum and its width (FWHM) a minimum at the transition, $T_c \approx 130K$. Fig.2 shows the relaxor behavior of the dielectric susceptibility. Its frequency dispersion becomes clearly visible below approximately 140K but is weaker than in other relaxors such as KLT. It is worth noting however that this frequency dispersion in KTN crystals with lower concentrations of niobium (e.g. 3%) is very similar to that in PMN and PZN although still weaker than in KLT.

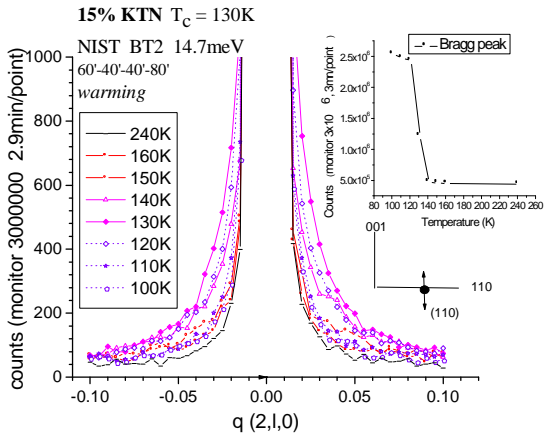

Fig.1a (110) Elastic diffuse scattering for different temperatures; inset: Bragg peak intensity vs *T*.

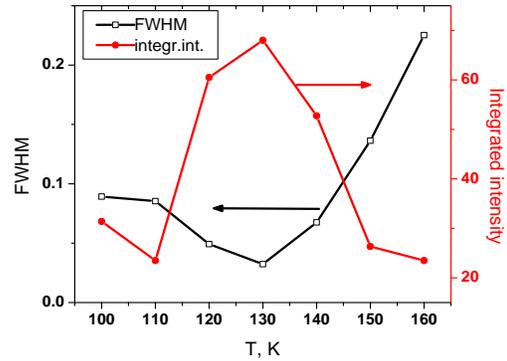

Fig.1b FWHM and integral intensity of the DS around (110) vs. *T*.

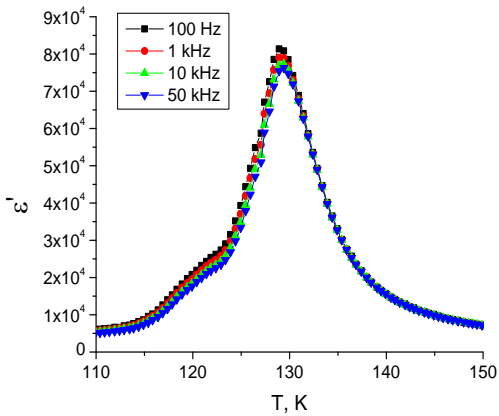

Fig.2. Dielectric susceptibility of KTN15 vs. temperature.

The minimum of the full width at half maximum (FWHM) of the diffuse scattering provides an estimate of the maximum correlation length at the transition, $\xi(T=130K) \approx 10$ unit cells $\approx 39.5$Å. Whether the diffuse scattering (DS) observed in KTN15 unequivocally indicates dynamic PNRs, the formation of static PNDs (within the energy resolution of the measurement) or both PNRs and PNDs is not a completely settled question. A likely answer to this question can however be obtained from the results of dielectric resonance studies by Toulouse *et al.* [14]. These resonances, which appear well above the transition, are due to polarization-acoustic strain (or piezoelectric) coupling in the paraelectric phase of KTN and other relaxors. Such a coupling should be absent in the perfect cubic phase of KTN and its existence reveals the presence of PNDs –or static correlations of Nb ion displacements in neighboring crystal cells- and the accompanying polarization and local strain fields. In a KTN15.7%Nb crystal for example, these resonances have been observed at temperatures as high as 200K upon cooling under an external dc electric field and up to approximately 180K upon zero-field heating, with a large change in amplitude around $T^* \approx T_c + 25$ K with $T_c \approx 139$K. [18] Their metastability indicates that the PNDs probably form as in a diffuse first order transition, similar to the condensation of water droplets in a supersaturated vapor. We should also note that, by analogy with a vapor, the presence of static PNDs does not preclude the simultaneous presence of PNRs with dynamically correlated Nb ions.

## 2.2 HIGH RESOLUTION INELASTIC NEUTRON SCATTERING in KTN15.

High resolution measurements of the transverse acoustic (TA) phonon in a single crystal KTN15% were also made on the SP4F2 cold neutron triple axis spectrometer at the Laboratoire Leon Brillouin in Saclay (France), near the (200) B.Z. point in the (100)-(010) scattering plane, with the phonon propagating in the (010) direction. Elastic diffuse neutron scattering (DS) was absent at that point [14], within the limit of our resolution, but was however present near the (110) BZ point, as reported above. For the inelastic neutron scattering measurements, the effective collimations used were, horizontally 273'-27'- 40'-40', vertically 51'-69'-137'-275', and the final neutron wave vector was 1.64Å$^{-1}$ from q=0.025 to q=0.9, providing a high energy resolution of ~0.2 meV (0.05THz), and 2.662Å$^{-1}$ at q=0.12 and 0.16, providing a resolution of 1meV. Pyrolytic graphite was also used to filter the harmonics. Several representative spectra are shown in Fig.3 for the TA phonon with $q$=0.025 and $q$=0.12 near the (200) Bragg reflection at different temperatures in the region where diffuse scattering is very small. Fitted curves are also shown, using a damped harmonic oscillator description for the phonon, including the thermal population, and taking into account the combined influence of the spectrometer resolution and of the strong anisotropy of the phonon dispersion surface away from the <100> direction, as indicated in the legend of Fig.3.

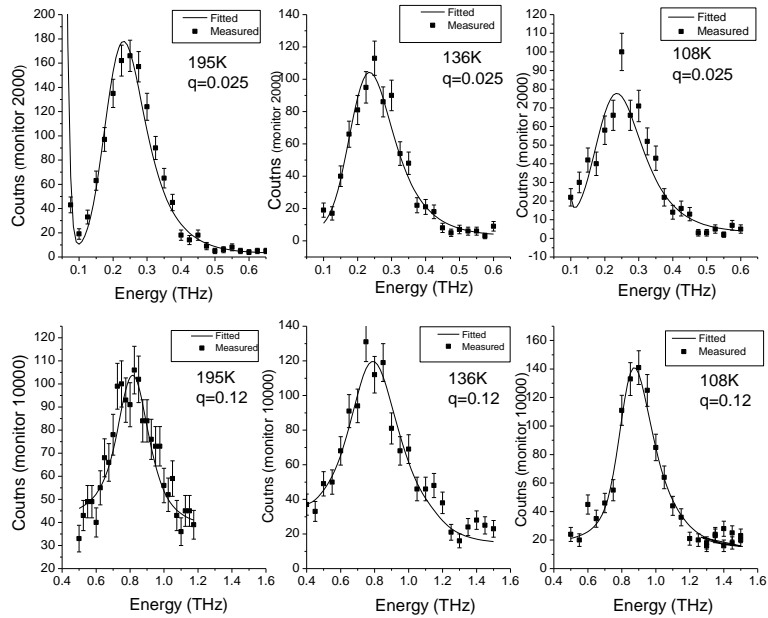

Fig.3 Transverse acoustic phonon spectra (experimental data and fitting curves for a damped harmonic oscillator). The functional form used to fit the spectra corresponding to q=0.025 was $\omega^2 = 26q_y^2 + 60q_x^2 + 30q_z^2$ and that for q=0.12, $\omega = 0.5 + 1.5q_y + 20q_x^2 + 20q_z^2$. The same functional form was used for all temperatures.

## TA PHONON DISPERSION

The TA phonon dispersion is presented as a function of wave vector in Fig. 4 and temperature in Fig. 5. As previously observed, the dispersion curve in Fig.4 exhibits a kink around $q$=0.12, still quite far from the zone boundary, which is reproduced here by a sine function. In Fig.5, the phonon frequency is relatively flat with temperature for small wave vectors, q≤0.07. But the frequency shows a clear dip around 136K (temperature measured) for q≥0.09 and a broad minimum also around the same temperature with a clear hardening at lower temperatures for q=0.16. As a rule, acoustic phonon frequencies increase at low temperatures due to anharmonic interactions. Here instead, a frequency minimum is observed (especially for $q$=0.16) in the region where strong polarization fluctuations are present (PNRs/PNDs).

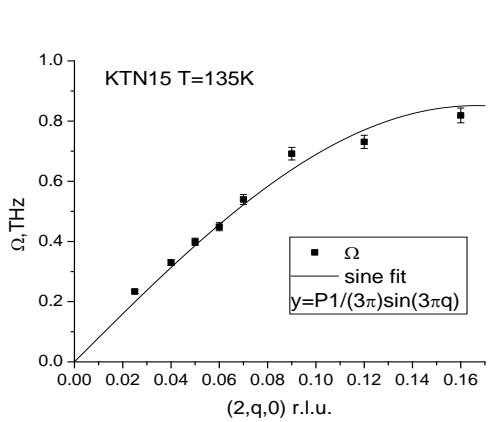

Fig.4 TAW dispersion curve for KTN15 at T=135K. Fitting curve $\Omega$=P1/(3π)sin(3π$q$) with P1=8.024±0.17 THz/r.l.u. Note the strong deviation (~14%) from a linear dispersion even at q~ 0.1.

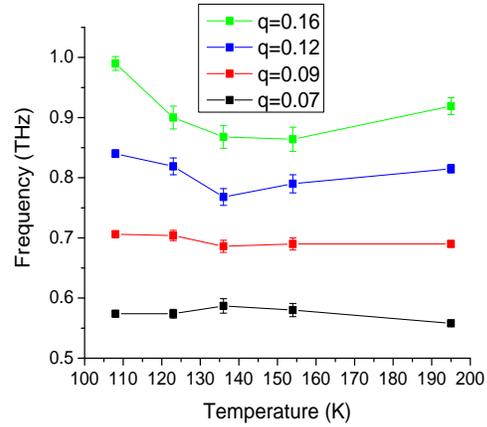

Fig.5. TAW (2,q,0): frequency vs. temperature for KTN15. (1THz=4.139 meV)

## TAW DAMPING IN KTN15

The evolution of the TAW damping as a function of temperature is shown in Figs. 6a-b.
In Fig.6a, the phonon width, $\Gamma$, is relatively flat and independent of temperature for small wave vectors. For larger wave vectors however, $\Gamma$ increases with decreasing temperature and reaches a maximum for $q$≥0.07 around 136K, a temperature at which the PNDs are clearly present (see diffuse scattering in Fig.1). $\Gamma$ then decreases significantly at lower temperatures. This temperature profile of the TA damping indicates that the lattice is relatively ordered at high temperature, becomes partially disordered at intermediate temperature, when the PNRs and then the PNDs are present and scatter the TA phonons, and returns progressively to a more homogeneous although still distorted state (large Bragg intensity) below $T_c$ as the PNDs grow and merge. It is also interesting to note that the curves corresponding to q≤0.08 all tend to converge toward ≈220K, the upper temperature limit at which the PNDs are known to appear (see dielectric resonances mentioned earlier [14] and again below).

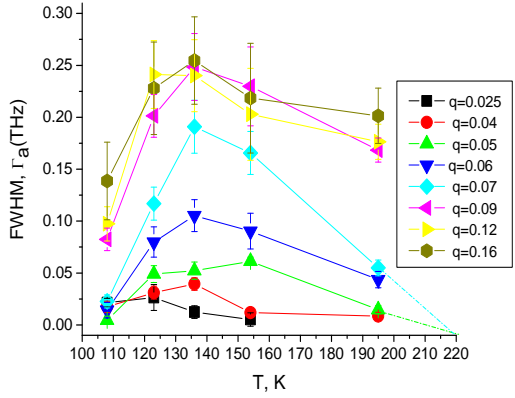 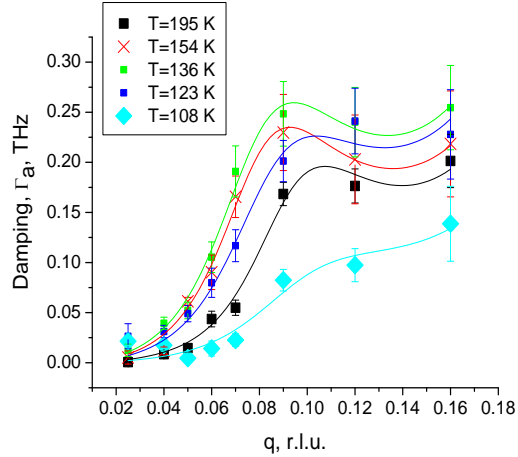

Fig.6a. Damping (FWHM), $\Gamma_a$, of the transverse acoustic (TA) mode vs temperature for different values $(2,q,0)$ in the single crystal KTN15. (1THz=4.139 meV, 1meV=11.59 K).

Fig.6b. TA phonon damping $\Gamma_a$ vs wave vector $q$ at the different temperatures. Experimental data and 3-parameter fit with Eq.(4) in the text.

In Fig.6b the $q$ dependence of the damping also reveals an interesting trend. At all temperatures, the phonon width exhibits a step increase around $q$=0.07, goes through a shallow maximum at $q$=0.09-0.12 and remains high beyond. The step increase is moderate at 195K, the highest temperature measured, maximum at intermediate temperatures around 140K and significantly smaller at the lowest temperature measured of 108K. Such a step increase with maximum around $q$=0.09-0.12 suggests scattering of the TA phonon by inhomogeneities of approximate size 8÷11 unit cells. As mentioned at the beginning of the paper, dynamic PNRs appear at Burns temperature,$T_d$ , then become quasi-static or static PNDs at $T^*$, giving rise to diffuse scattering (DS) and the characteristic frequency dispersion of the dielectric constant of relaxors. As seen in Fig. 2, $T^*$≥160K in KTN15. In fact, based on the observation of the dielectric resonances mentioned earlier, we conclude that PNDs probably become stable at temperatures as high as T=200K in KTN15.7 upon cooling in an external electric field.[14] It is important to note that the temperature at which the PNDs are observed to appear also depends upon the characteristic frequency of the measurement, that temperature being higher for higher frequencies.

2.3 INELASTIC NEUTRON SCATTERING in KTN17.

To help with the interpretation of the high resolution inelastic neutron scattering data presented here for KTN15%, additional data on the TA and soft TO phonon mode dynamics were obtained on another single crystal with a similar concentration, KTN17 (17%Nb), which exhibited similar features to those in KTN15. Measurements of KTN17 were made on the thermal neutron triple axis spectrometer BT9 at the NCNR neutron center (NIST) with fixed incident energy of 14.7 meV [21]. Constant $q$ and constant $E$ scans of the TO and TA phonon branches were performed along the line ($q$, 2, 0) in the temperature range between 100K and 310K. The dispersion curves obtained directly from the observed frequencies in the inelastic spectra measured at constant $q$ and at higher temperatures (310K and 240K) are shown in Fig.7a.

The soft TO phonon peaks are seen to be already unusually broad at these higher temperatures as indicated by the large vertical bars. The temperature dependence of the TO phonon frequency at the (0,2,0) B.Z. point is shown in Fig.7b. The measured spectra are presented below in terms of the dynamic structural factor, $|F(Q)|$, which is proportional to the scattering cross section and is obtained from the experimental intensities, $I$, after correction for the Bose temperature factor and assuming an inverse square root energy dependence of the harmonic phonon displacement amplitude (see Shirane *et al*. [20a]):

$$I = A \frac{1}{E(q)} |F(Q)|^2 \begin{Bmatrix} n+1 \\ n \end{Bmatrix} \text{ with } n = \frac{1}{\exp[E(q)/T]-1} \qquad (1)$$

in which $E$ is the phonon energy and $A$ is a numerical factor determined by the parameters of the spectrometer. We note that the transformation given in Eq. (1) corrects for the varying scattering weight at different energies, but unfortunately also enhances background fluctuations. We also note that effects of the spectrometer resolution are not taken into account in such a simple approach. In the following we will omit the factor $A$ for the purpose of a semi quantitative analysis.

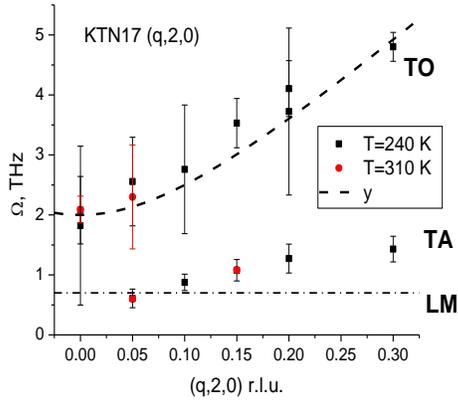
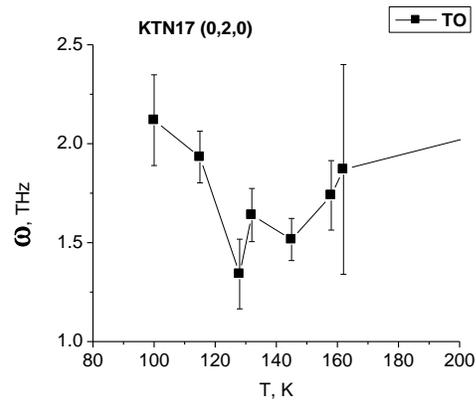

Fig.7a KTN17: dispersion of the TOW and TAW at T=310K and 240 K. Note the large width of the TO peak (vertical bars).
$y=\sqrt{\omega_0^2+C_o^2 q^2}$, $\omega_0$=2 THz, $C_o$=15 THz/r.l.u.
The dot line at E=2.9meV marks the estimated energy of the dispersionless localized modes mentioned in the text.

Fig.7b KTN17. Temperature dependence of the TO phonon frequency at the (0,2,0) B.Z. point. TO phonon is softening but its frequency remains high even at zone center with a gap $b$~1.2 THz. The vertical bars indicate the phonon peak width.

Constant energy spectra measured at $E$=4meV and 6meV are reproduced in Figs.8 and 9. These two energies were conveniently chosen to separately probe the behaviors of the TAW (4meV) and TOW (6meV). As seen in Fig.8a for $E$=4meV, the overall TA phonon intensity is smaller in the temperature range, 130K<T<160K with significant intensity present at the zone center, $q$=0. $F^2$ returns to a more usual phonon spectral shape at 100 K. The same TA phonon curves are presented in a more detailed manner and fitted with damped harmonic oscillator functions in Figs.8b-c at T=130K and 100K respectively. Note that the different peak shapes for positive and negative *q* are only an effect of the spectrometer resolution (focusing for positive *q* and defocusing for negative *q*).

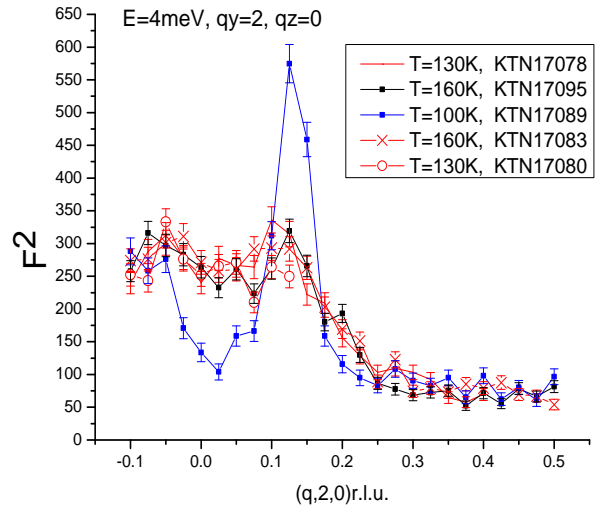

Fig. 8a KTN17: *q*-scan at constant energy, for several temperatures upon cooling and warming (**E**=4 meV=0.966 THz).

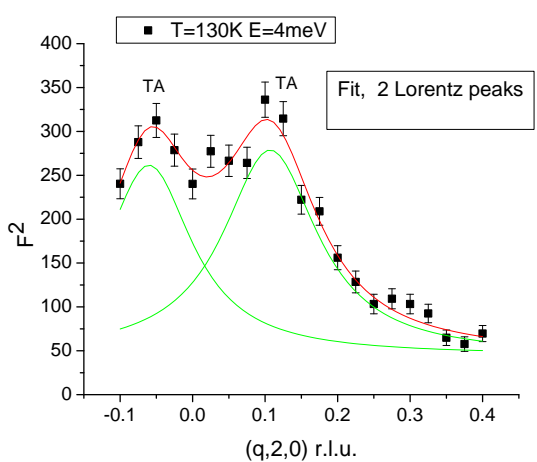

Fig.8b KTN17. T=130K. *q*-scan at constant energy, E=4 meV. Fit:2 Lorentz peaks, positions $q_{c1}=0.107$, $q_{c2}=-0.06$; widths $w_1=0.158$, $w_2=0.144$; areas $A_1=58$, $A_2=49$.

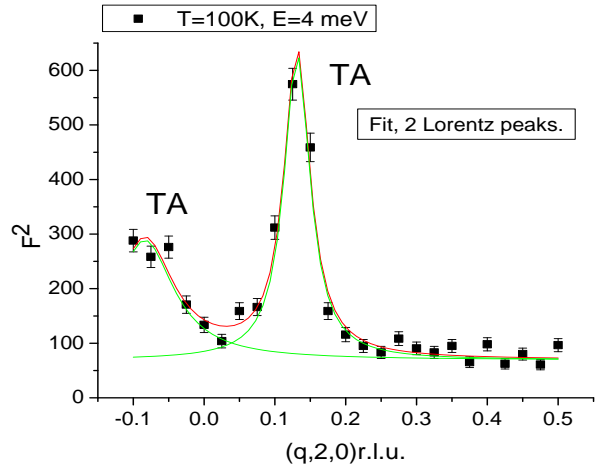

Fig.8c KTN17. T=100K. *q*-scan at constant energy, E=4 meV. Fit:2 Lorentz peaks, position $q_{c1}=-0.084$, $q_{c2}=0.131$; width $w_1=0.1$, $w_2=0.045$; area $A_1=34.5$, $A_2=39.5$

Constant E spectra taken at E=6 meV and fitted peaks are shown in Fig.9. At this energy, only one peak corresponding to the TO phonon is observed in the spectra, consistent with the dispersion curves shown in Figs.7a-b. (note: the two peaks at T=130K correspond to the same phonon at positive and negative $q$) The width of the TO phonon peak is seen to increase or equivalently the correlation length of the TO phonon to decrease with temperature. The high background at 130K and 100K points to a very broad distribution of intensity in $q$-space corresponding to a very short correlation length, or localized excitations (LM). In Fig. 9c at 100K, the TO peak is absent.

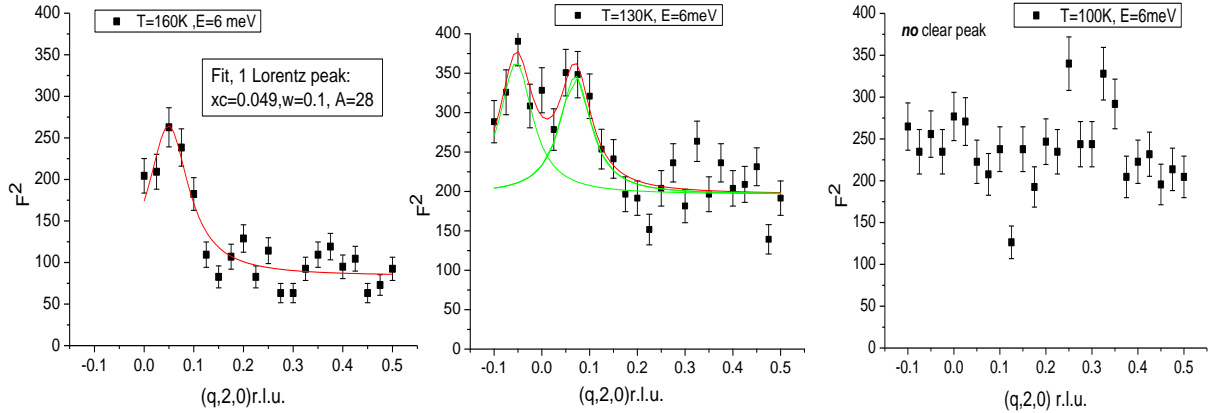

Fig.9 KTN17; $q$-scan of the TO phonon at constant energy, E=6 meV at a) T=160 K, the correlation length is L=1/w=10 l.u. ≈ 40Å; b) T=130 K, two Lorentz peaks are seen on higher background; c) T=100 K, no peak but only a high background

In trying to understand the origin of the TA phonon damping in KTN15, the main experimental results obtained on the KTN15 and KTN17 single crystals are the following:
  - a deviation from the linear dependence in the TA dispersion curve, ω vs $q$, between q≈0.1 and 0.16 and a broad minimum in the TA frequency around the transition temperature, T=135K, in this same q range.
- a step-like increase in TAW damping with inflection point around $q$=0.07
- maximum TA damping at T=135K for $q$ between 0.12 and 0.16
- a high minimum of the TO mode frequency near the transition
- no observed crossing of the TA and soft TO phonon branches, at least for $q \leq 0.16$, but crossing of the TA phonon branch and a dispersionless branch of localized excitations (LM).

In the following analysis, we show that the high frequency of the soft TO mode in the transition region, T<160K, is due to its strong coupling to localized excitations (LM) whose flat branch at $\omega_L$=0.7 THz crosses the TA curve. These localized excitations can be attributed to the local reorientational motion of the off-center Nb ions through tunneling. The TA damping is then due to scattering of the TAW by mixed TO-LM excitations.

## 3. ANALYSIS of RESULTS

### 3.1. Theoretical analysis of the TAW resonant damping.

The central experimental facts that we seek to explain is the TA phonon broadening with decreasing temperature, its maximum damping in the temperature range in which PNDs are present and its return to low damping at low temperatures, despite little or no diffuse scattering near the (200) reflection in KTN [8]. The model presented below and in the appendices to explain these experimental results is inspired by the model proposed by Axe et al [15] who considered the effect of a direct bilinear TA-TO interaction on the TA dispersion in the incipient ferroelectric $KTaO_3$. Here we follow a similar approach, but additionally take into account the mixing of the TO and LM modes on the one hand and the local polarization $P$ that develops with PNRs/PNDs on the other. We thus write an interaction Hamiltonian, $H_{fr}$:

$$H_{fr} = 2f(P\xi)_{ik} u_{ik}, \quad (P\xi)_{ik} = 1/2(P_i\xi_k + P_k\xi_i - 2/3\delta_{ik}(P\xi)),$$

$$u_{ik} = 1/2(\partial u_i/\partial r_k + \partial u_k/\partial r_i - 2/3\delta_{lk}\text{div}(u)) \qquad (2)$$

in which $\xi_k$ and $u_k$ are the $k$- components of the TO and acoustic (shear) mode displacements respectively, and the parameter $f$ is the coupling constant. As shown in Appendix B (B8), the LM modes, which are associated with transitions between $Nb^{5+}$ tunneling states, can only become excited by the TA phonon with the appearance of the quasi-static or static polarization $P$ of PNRs or PNDs. The similarity of Eq.(2) with that used in Axe et al.[15] is not a direct but only a formal one because $\xi_k$ here represents a component of the renormalized TO mode eigenvector, taking into account the strong TO-LM interaction mentioned earlier. In the following, TO and LM refer to the renormalized modes resulting from this interaction. The normally soft TO mode is renormalized toward higher frequencies and the LM modes in our case toward lower frequencies where they interact with the TA phonon, causing its observed damping. We work here in the approximation of a homogeneous medium and neglect in first order approximation any dispersion of the resonance. Also, unlike the TO-LM interaction, we suppose that the interaction between the TAW and the mixed LM modes is not strong (weak coupling approximation). Assuming the simplest form of a plane TA wave propagating along the $z$ axis, polarized along the $x$ axis and interacting with the mixed LM wave polarized along the same $x$-axis, the Lagrangian $\Lambda$ and dissipative function $\Psi$ can be written as:

$$\Lambda = \int L d^3 r, \quad L = 1/2(\partial u/\partial t)^2 - C_a^2/2(\partial u/\partial z)^2 + 1/2(\partial \xi/\partial t)^2 - \omega_L^2/2\xi^2 - fP_z\xi \partial u/\partial z,$$

$$\Psi = 1/2\int \Gamma_L (\partial \xi/\partial t)^2 d^3 r \qquad (3a)$$

in which $C_a$ is the velocity of sound. The coupled equations of motion for the TA and LM mode with momentum $q$ and frequency $\omega$ can then be written as:

$$(\omega^2 - \omega_a(q)^2)u + ifqP_z\xi = 0, \quad \text{with} \quad \omega_a(q) = C_a q$$

$$-ifqP_z u + (\omega^2 - \omega_L^2 + i\omega\Gamma_L)\xi = 0 \qquad (3b)$$

in which $\omega_a(q)$ and $\omega_L$ are the frequencies of the TA and localized mode respectively, $\Gamma_L$ is the damping of the localized mode, $P$ the polarization from PNRs/PNDs and $f$ is the coupling constant. Solving these coupled equations using perturbation theory to the first non-vanishing order in $f$, we obtain the following expression for the acoustic damping:

$$\Gamma_a(q) \approx M^2 q^2 \frac{\Gamma_L}{(\omega_a(q)^2 - \omega_L^2)^2 + \omega_a(q)^2 \Gamma_L^2} \quad \text{with} \quad M^2 = f^2 P_z^2/2 \qquad (4)$$

It is interesting to note that this damping is proportional to the square of the oscillator strength, $M$, itself proportional to the local polarization, $P$, from PNRs/PNDs. The experimental TA damping data shown in Fig.6b are fitted with expression (4), taking into account the sine approximation shown in Fig.4 for the TA mode dispersion and considering $\omega_L$, $\Gamma_L$ and $M$ as fitting parameters. The values of the fitted parameters are presented in Figs.10a and b. The temperature dependencies of the frequency and width of the resonance are seen to be moderate. By contrast, its oscillator strength, $M$, exhibits a strong temperature dependence ($M^2$(T=136K)/$M^2$(T=108K)~2.7), which mirrors the experimentally determined TA damping shown at Fig.11b. The maximum of the TA phonon damping can then be attributed to the resonance maximum of the TA-LM coupling which is discussed below.

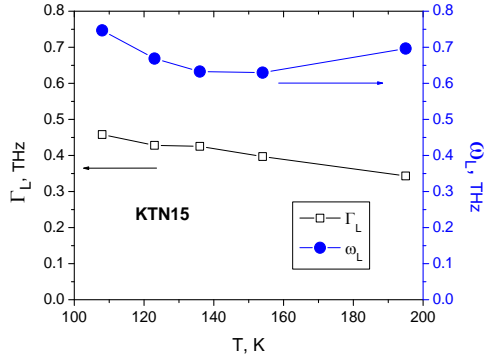

Fig.10a. Frequency and width of the localized mode vs temperature, obtained by fitting the experimental data in Fig.6b; note that $\Gamma_L < \omega_L$ at all temperatures.

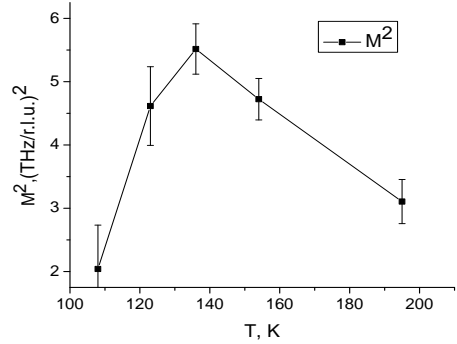

Fig.10b. Oscillator strength $M^2$ vs temperature.

## 3.2. TAW damping as due to coupling to a relaxation.

Alternatively, the temperature and wave vector dependence of the TA damping can be described phenomenologically as due to the direct coupling between the TAW and a purely relaxational local mode, distinct from the resonance mode described in expression (4). In the framework of the standard Debye relaxation model, the TAW damping would be written as follows:

$$\Gamma_R(q) = qG_D \frac{\omega(q)\tau_D}{1+[\omega(q)\tau_D]^2} \qquad (5)$$

Here $\omega$ is the frequency of the TAW, $\tau_D$ designates the relaxation time of the local mode and $G_D$ is a numerical factor. The dispersion, $\omega(q)$ of the TAW is taken from the experimental data shown in Figs.4-5. The TA damping curves are now fitted with the two free parameters, $\tau_D$ and $G_D$. However, such a simple Debye-type relaxation is found to be unable to reproduce the step-like increase observed in the measured TAW damping around $q$=0.07. Instead, the presence of PNRs/PNDs suggests a distribution of relaxation times. Under this assumption, the TA phonon damping can be described in the framework of the Havriliak-Negami relaxation model for the permittivity, $\chi(\omega)$, (see [11], [15]), which is an empirical modification of the Debye model with two exponents $\alpha$ and $\beta$. The exponent $\alpha$ is a measure of the distribution of relaxation times and $\beta$ introduces an asymmetry in the distribution ($\beta$=1 for a symmetric distribution):

$$\chi(\omega) = \chi_\infty + \frac{\Delta\chi}{[1+(i\omega\tau)^\alpha]^\beta} \quad \text{with} \quad \Delta\chi = \chi_S - \chi_\infty \tag{6}$$

in which $\chi_\infty$ is the high frequency limit of the permittivity, $\chi_S$ the static low frequency permittivity and $\tau$, a characteristic average relaxation time. Special cases of the HN relaxation function correspond to the Debye ($\alpha=\beta=1$), Cole-Cole ($0<\alpha<1$, $\beta=1$), and Cole-Davidson ($\alpha=1$, $\beta\neq1$) processes. The Cole-Cole relaxation is often used to describe the so-called stretched relaxation observed in glasses and polymers. As shown in Fig. 11a and b, using as before the sine approximation to the TA phonon dispersion, $\omega(q)$, shown in Fig.4, the TA phonon damping can be fitted with excellent accuracy by the imaginary part of the susceptibility expression (6), with $\beta=1$ and three free parameters, $P$, $\tau$ and $\alpha$ with $\alpha\approx3/2$:

$$\Gamma_a = qP \frac{[\omega(q)\tau]^\alpha \sin(\pi\alpha/2)}{1+2[\omega(q)\tau]^\alpha \cos(\pi\alpha/2)+[\omega(q)\tau]^{2\alpha}} \tag{7}$$

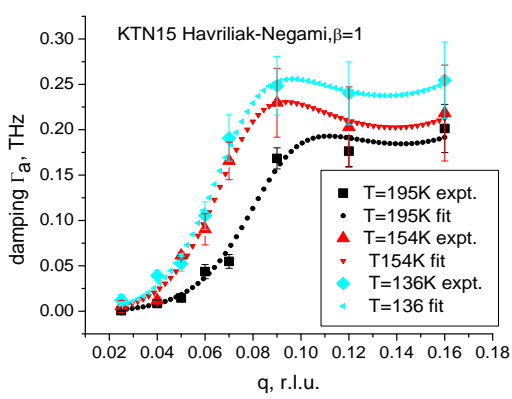
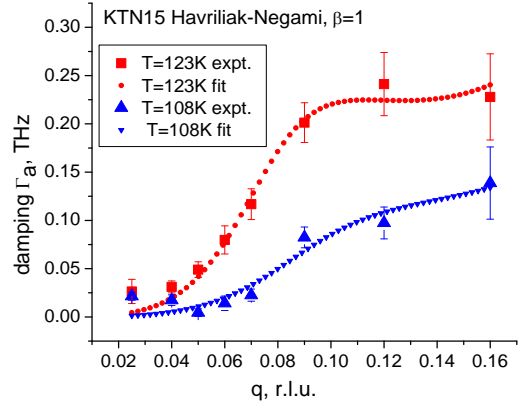

Fig.11a Experimental data and fit with expression (6) for the Havriliak-Negami relaxation model with $\beta=1$ at T=195K, 154K, and 136 K.

Fig.11b Same as in Fig.12a but at T=123K and 108K. A step-like increase in the TAW damping around $q$=0.07 is precisely reproduced in this H-N model.

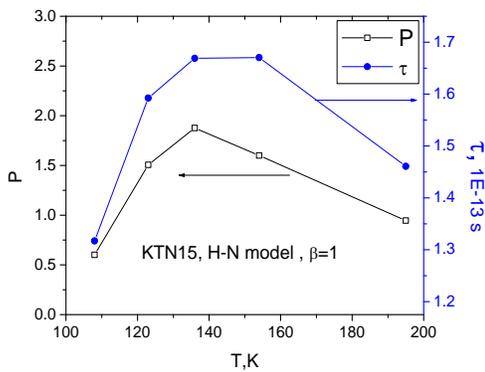
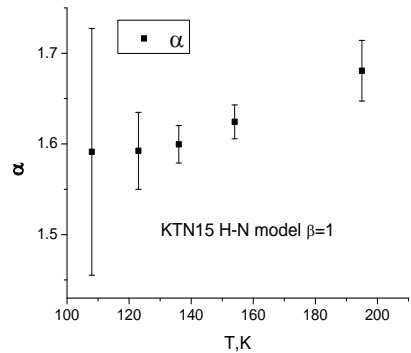

Fig.11c Values of the fitting parameters $P$ and $\tau$ vs. temperature using the Havriliak - Negami relaxation model in Eq. (7) with $\beta=1$.

Fig.11d Value of the fitting exponent $\alpha$ vs. temperature using the Havriliak - Negami relaxation model in Eq. (7) with $\beta=1$.

The fitting parameters are plotted in Fig.11c and d.

We now discuss specific features of the HN function connected with the value of the exponent $\alpha$. Comparing the Debye, $f_D$, and Havriliak-Negami, $f_{HN}(\beta=1)$ relaxation functions:

$$f_D = 1/(1+i\omega\tau) \quad \text{and} \quad f_{HN} = 1/(1+(i\omega\tau)^\alpha) \tag{8}$$

We note that $f_{HN}$ with $\alpha=1/2$ (i.e. <1) slowly decreases for $\omega\tau\gg1$, which is often described as a stretched relaxation in the literature of glasses and polymers. By contrast, the same function $f_{HN}$ but with $\alpha\approx3/2$ (i.e.>1), as in the present case, gives rise to a resonance-like transient oscillation that is responsible for the step-like increase in the TAW damping observed experimentally around $q=0.07$.

In the time-domain, the Debye relaxation function monotonically drops [25] to zero with time:

$$F_D(t) \equiv \frac{1}{2\pi}\int_{-\infty}^{\infty}\frac{\exp(i\omega t)}{1+i\omega\tau}d\omega, \quad F_D(t<0)=0 \quad \text{and} \quad F_D(t>0) = \frac{1}{\tau}\exp(-t/\tau) \tag{9}$$

The Havriliak-Negami relaxation function with $\alpha>1$ exhibits a resonance-like oscillation at short times. Its time-domain representation, $F_{HN}(t)$, is shown in Fig.12 and can be expressed by means of the Mittag-Leffler function $E_\alpha(x)$ for the case $\beta=1$ [16, 17].

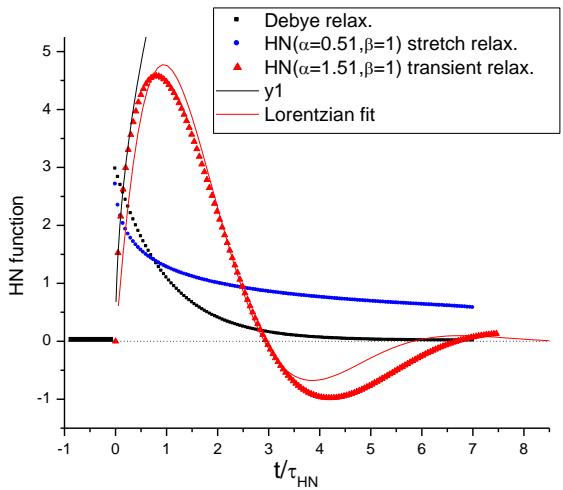

Fig.12 Time evolution of the Debye, stretched HN (with α=0.51, β=1) and transient HN relaxation (with α=1.51, β=1), and Lorentzian fit y=A*exp(-t/τ/t₀)sin(π t/τ/w) with A=10.56, t₀=1.49, w=2.92. For t/τ$_{HN}$ <0.35, the HN function can be approximated by the expression y₁=6.7838(t/τ$_{HN}$)$^{1/2}$. For convenience the Debye and stretched relaxation curves are displaced vertically.

The resonance-like transient relaxation indicates that the Havriliak-Negami function $f_{HN}$ ($\alpha=3/2$, $\beta=1$) (8) has a pole in the complex $\omega$ – plane (see Appendix A for details) and can also be adequately fitted at short times by a Lorentzian function as illustrated in Fig.13. Another significant difference between the $f_D$ and $f_{HN}$ relaxation functions is worth mentioning. A Debye relaxation is a localized process in which each unit relaxes independently of the others after several rapid initial collisions. In the high-frequency range, $\omega\tau\gg1$, or at short times, $f_D$ ($\alpha=1$) can be expanded in terms of integral powers of the parameter $1/(\omega\tau)$ while $f_{HN}(\alpha=1/2)$ is expanded in terms of half powers, $1/(\omega\tau)^{1/2}$. This point was carefully investigated in the context of high-viscosity liquids. A detailed theory was proposed by Isakovich and Chaban [18] who regarded these liquids as micro-inhomogeneous media whose dynamics are controlled by diffusion, as in a delocalized or non-local process. Letting $R$ designate the characteristic size of a micro-formation or a diffusion length in the viscous liquid, the long wave response of such a medium to an excitation with wave vector $q$ will be defined by the value of the dimensionless parameter $(qR)$, or equivalently by the value of $(\omega\tau)^{1/2}$. As discussed in the second part of the discussion section below, $\alpha\approx3/2$ indicates that the local relaxation of the polarization has now become slower than its diffusion time.

## 4. DISCUSSION

We have studied the damping $\Gamma_a$ of the TA phonon in the relaxor KTN with moderate niobium concentrations of 15%-17% by means of high resolution inelastic cold neutron scattering. The TA phonon linewidth or damping exhibits a step increase around momentum $q$=0.07, goes through a shallow maximum at $q$=0.09-0.12 and remains high beyond and up to the highest momentum studied of $q$=0.16. This step increase is moderate at 195K, the highest temperature measured, maximum at intermediate temperatures around 140K and significantly smaller at the lowest temperature measured of 108K.

The observed TAW damping is successfully fitted in the framework of the simple model described above of the TA phonon interacting with a mixed LM-TO mode. From fitting the data in Fig.6b using Eq.4, the bare localized mode (LM) is found to have frequency $\omega_L$ ~0.6÷0.7 THz and damping $\Gamma_L \leq \omega_L$. Maximum scattering of the TAW occurs when the frequency of the acoustic phonon is resonant with the localized mode, $\omega_a \approx \omega_L$. As seen in Fig.10a, the bare frequency of the LM mode, $\omega_L$, depends only moderately on temperature (variation ≤ 15%) and the variation of $\Gamma_L$ with temperature is less than 30%. However, the oscillator strength $M$ is seen in Fig. 10b to exhibit a strong maximum in the temperature range ~120K÷160 K, which is also the temperature range of the dielectric relaxation maximum (relaxor behavior).

We now address the possible nature of the resonant local mode (LM) in KTN. As suggested earlier in the present paper, the existence of such a mode is connected with the presence of off-center $Nb^{+5}$ ions, displaced from their high symmetry sites by 0.145 Å in one of eight equivalent <111> directions inside PNRs/PNDs [9]. The off-center Nb ions undergo two types of dynamics, on two different time scales: the first one is associated with the TO mode and the second one with the tunneling motion of the $Nb^{+5}$ off-center ions within each unit cell. Whereas the intrinsic soft TO branch would normally tend toward zero at the center of the Brillouin zone, here the observed TO phonon frequency remains high even at zone center with a gap $b$~1.2 THz (Fig.7b). The presence of a similar gap, albeit smaller, was also reported in $SrTiO_3$ [28], $\omega^c_\infty \approx 0.13$ THz, and attributed to the interaction between the TO phonon and an unspecified relaxation mode (central peak theory). The existence of the gap $b$ in KTN could similarly be attributed to the presence of an additional spectral density in the TO-TO phonon correlation function, similar to the CP in $SrTiO_3$. However, the much larger gap in KTN, $b$~1.2 THz near the (020) B.Z point, suggests an additional coupling between the TO phonon and excitations with energy ~0.7 THz, such as those attributed to the reorientation tunneling motion of the $Nb^{5+}$ off-center ions (LM).

### *Pseudo-spin model of the TA-[TO-LM] interaction*

The model discussed below rests on the assumption that the $Nb^{+5}$ ions are likely involved in both the acoustic TA and the soft optic TO motion as well as in their own tunneling motion and/or thermally activated jumping between equivalent sites. Girshberg and Yacoby [10, 18] have described the dynamics of a system of off-center ions in terms of pseudo-spins. A similar formal approach has been applied to order-disorder ferroelectric transitions. [20] Pseudo-spin dynamics is usually described in the framework of the Ising model with a spin Hamiltonian that also includes a transverse field, $H_T$: [18]

$$H = -\Omega_0 \sum_{\mathbf{m}} \sigma^x_{\mathbf{m}} - \frac{1}{2} \sum_{\mathbf{m},\mathbf{m}'} J(\mathbf{m}-\mathbf{m}') \sigma^z_{\mathbf{m}} \sigma^z_{\mathbf{m}'} + H_T \qquad (10)$$

Here $m$ is the position index, $\sigma_m$ is the Pauli spin matrix, $J(m-m')$ is the direct spin-spin interaction constant, and $\Omega_0$ is the tunneling frequency of an individual off-center ion between different orientations. The term $H_T$ -'the heat reservoir' – includes the interaction of the spins with all thermal excitations, namely thermal phonons, but excluding the TO phonon which is being considered separately. The strength of the direct spin-spin interaction $J(m-m')$ is assumed to be moderate. Nevertheless, spin-spin interactions through a virtual TO mode are essential and become particularly strong near the transition. In KTN, Girshberg and Yacoby [10, 18] estimated the value of the individual $Nb^{+5}$ ion tunneling splitting between off-center positions to be $2\Omega_0 \sim 30 \div 34$ K$=0.625 \div 0.709$ THz. This is almost exactly the value $\omega_L$ obtained in Fig.6b from fitting the TA damping and also shown in Fig. 10a as a function of temperature.

We now formally describe the nature of the interaction between the TA and the TO phonon coupled to the local mode (LM). Because of the small off-center displacement, the direct electric dipole-dipole interaction between $Nb^{+5}$ off-center ions can be considered as moderate and we can therefore limit ourselves to the dynamics of independent off-center Nb ions tunneling between equivalent (111) positions (see Appendix B). Due to tunneling, the initially degenerate ground state splits into eight levels: a new completely symmetric ground state with energy $E=-3\Omega$, a dipolar triplet level ($E=-\Omega$), a quadrupolar triplet level ($E=+\Omega$), and an upper singlet state ($E=+3\Omega$). $Nb^{+5}$ ions are off-centered from their high symmetric crystallographic position by $R_{Nb}=0.145$ Å in a (111) direction and we can therefore expand the interaction of the TO and TA phonons with the tunneling $Nb^{+5}$ ions in Appendix B in terms of the small dimensionless parameter $r_{Nb} \equiv R_{Nb}/l.u.=0.036<<1$. (*l.u.*: lattice unit). The first order term in the expansion, $V^{(1)}(TO/SR) \sim r_{Nb}$, describing the TO phonon-off-center Nb interaction (B4), excites dipolar transitions with energy $2\Omega_0 \approx 0.6 \div 0.7$ THz and leads – in combining with the anharmonic TO-TO interaction -to the opening of the gap $b$. The second order term, $V^{(2)}(TO/SR) \sim r_{Nb}^2$, excites quadrupolar transitions with energy $4\Omega_0 \approx 1.2 \div 1.4$ THz and also contributes to the formation of the gap $b$ in Fig.7b. Because $r_{Nb}<<1$, it is reasonable to expect that the first order term (dipolar transition) dominates. In addition, equations B8-B10 show that the TA phonon will excite transitions with energy $2\Omega$ only if the system is at least partially polarized, i.e. in the presence of PNRs/PNDs. We therefore can expect that the TA phonon damping (at $2\Omega=\omega_L \approx 0.6 \div 0.7$ THz, Fig. 11a-b) will be small at high temperature in the absence of PNRs/PNDs, maximum at $T \sim 140$K near the transition in the presence of PNRs/PNDs and decreasing at low temperature. The TA phonon could also excite transitions with energy $4\Omega$ due to the interaction $V(TA/SR) \sim r_{Nb}^2$ of an elastic dipole (*tensor*) with acoustic stress (B6), (B7). However such high-energy TA phonons are not observed in our experiment (Fig.8). A detailed analysis of the TO-LM interaction effect on the genesis of the PNRs/PNDs in KTN is outside the scope of the present paper and will be published subsequently.

*TA damping and Diffuse Scattering*

We finally return to the question of the correlation between the TAW damping and diffuse scattering (DS). Stock et al [5] pointed out the existence of a correlation between their DS scattering results and the TAW damping in the relaxor PMN. These authors showed that strong DS near the (110) B.Z. point was accompanied by heavy damping and even overdamping of the TA phonon. Based on this result, they concluded to a firm connection between DS and TA damping at a particular BZ point. Recently however, Stock et al. [6] have also reported in a PMN single crystal a moderately damped TA phonon ($\Gamma/\omega \sim 0.3 \div 0.4$ at $q=0.21$) near the (200) reflection where DS is small or absent, as in our KTN results.

Therefore, their results and ours may not be in contradiction with each other and the connection between the DS and the TAW damping is not a necessary one but may depend on the particular situation.

We noted in the Introduction that Hirota et al. [4] explained their neutron diffuse scattering data in PMN by proposing that diffuse scattering is not necessarily due to a condensed soft TO mode but can also be due to a collective shift **δ** of all the atoms within a PNR along the local polar direction. What can be said of DS in KTN in light of the above discussion? Is there a δ-shift in KTN as in PMN [4]?

One of us has previously shown that a localized mode can be described as a *coherent* superposition of the TO and TA phonons [24], which is similar to the collective shift *δ* along the polarization direction proposed by Hirota *et al*. [4]. However, such a shift is only possible in the case of a strong TO-TA interaction. Although there is presently no definite information regarding the strength of the TA-TO interaction in KTN, recent observations of characteristic electro-mechanical resonances in several relaxor ferroelectrics by Toulouse *et al*. [13, 24] ($K_{1-x}Li_xTaO_3$ (KLT) and $KTa_{1-x}Nb_xO_3$ (KTN) and $PbZn_{1/3}Nb_{2/3}O_3$ (PZN)) reveal the piezoelectric character of the polar nanodomains (PNDs). These resonances are macroscopic manifestations of a relatively strong TO-TA coupling within PNDs. Here, we have shown that this interaction also involves the local modes (LM) associated with the tunneling of the off-center Nb. The results presented here should be useful for the study of other relaxor ferroelectrics with homovalent cations $(K_{1-x}Li_x)TiO_3$ (KLT), $Ba(Zr_{1-x}Ti_x)O_3$ (BZT), and $Ba(Sn_{1-x}Ti_x)O_3$ (BST) in which strong random fields will be absent. In the lead relaxors PMN and PZN, the heterovalent substituted ions give rise to random electric fields (REFs) that should be somewhat screened by the off-center ions and may also partially suppress the tunneling motion of the Nb ions within individual unit cells.

5. CONCLUSION

In the present paper, we have reported measurements of the transverse acoustic (TA) and transverse optic (TO) phonon in KTN crystals with Nb concentrations of 15% and 17%. In addition to the TO and TA phonons, these results suggest the existence of localized modes (LM) which can be associated with the tunneling reorientation of the off-center $Nb^{5+}$ ions. These LM modes are strongly coupled with the TO phonon and interact with the TA, giving rise to increased damping for intermediate values of the wavevector, $q \geq 0.07$, which corresponds approximately to the size of the polar nanodomains (PNDs). Transitions between $Nb^{5+}$ tunneling states that would otherwise not be excited by the TA phonon become allowed with the appearance of the quasi-static polarization of PNDs. The TO-LM coupling also qualitatively explains the incomplete softening of the TO mode. Conversely, it is the displacements of the same Nb ions correlated through the TO field that give rise to the formation of PNRs and PNDs. Therefore these Nb ions take part simultaneously in both the relatively high frequency TO motion and the relatively slow off-center or pseudo-spin correlated motion, i.e. a two-time scale dynamics [26]. At high temperature, the TO mode-driven Nb dynamics and the Nb off-center pseudo-spin dynamics are almost independent from each other because of their large difference in frequency or time scales and the displacements of different off-center Nb ions are only moderately correlated. Upon cooling, the evolution of the system can be described as follows: i) the pseudo spin-pseudo spin interaction strengthens and the displacements of the off-center Nb ions become correlated; ii) the dynamics of the TO and pseudo-spin becomes mixed and a coupled TO-pseudo spin mode appear; iii) this mixed LM mode eventually condenses resulting in the appearance of quasi-static or static atomic displacements and local polarization *P* within regions or "droplets" (PNDs), local strain fields, the relaxor behavior of the dielectric susceptibility and the dielectric resonances mentioned earlier [13,24]. The TA phonon damping reaches a maximum at ~136K (Figs.6b and 10b) due to the increasing number of PNDs. At low temperature, the coupling between the TO mode and the off-center Nb pseudo-spin dynamics weakens because the reorientation frequency of the off-center Nb falls away from the increasing soft mode frequency and the crystal returns to a regime similar to that of "ordinary" ferroelectrics. In such a scenario, the THz frequency TA phonon excited by neutrons acts as a probe of the TO – $Nb^{+5}$ dynamics in KTN.

ACKNOWLEDGMENTS

We thank Dr. P.Gehring for useful discussions. The experimental part of this work was initially supported by the US Department of Energy under grant DE-FG-06ER46318.

APPENDIX A

The Havriliak - Negami relaxation model

The Havriliak-Negami relaxation [11], [15], [17] is an empirical modification of the Debye relaxation model that takes into account the asymmetry and breadth of the dielectric dispersion curve by adding two exponential parameters $\alpha$ and $\beta$ to the Debye equation (6-8). The asymmetry and breadth of the corresponding spectra are described by the parameters $\alpha$ and $\beta$. The Debye relaxation correspond to the case $\alpha=1, \beta=1$.

The HN model can be described as a superposition of Debye models with a continuous distribution $g(\ln\tau_D)$ of the relaxation time, $\tau_D$:

$$\frac{\chi(\omega|\alpha,\beta) - \chi_\infty}{\Delta\chi} = \int_{\tau_D=0}^{\infty} \frac{1}{1+i\omega\tau_D} g(\ln\tau_D) d\ln\tau_D,$$

$$g(\ln\tau_D|\alpha,\beta) = \frac{1}{\pi} \frac{(\tau_D/\tau)^{\alpha\beta} \sin(\beta\theta)}{((\tau_D/\tau)^{2\alpha} + 2(\tau_D/\tau)^\alpha \cos(\pi\alpha) + 1)^{\beta/2}},$$

$$\theta = \arctan\left(\frac{\sin(\pi\alpha)}{(\tau_D/\tau)^\alpha + \cos(\pi\alpha)}\right) \text{ if the argument of the arctangent is positive, else}$$

$$\theta = \arctan\left(\frac{\sin(\pi\alpha)}{(\tau_D/\tau)^\alpha + \cos(\pi\alpha)}\right) + \pi \quad (A1)$$

The HN model (with $0<\alpha<1, \beta=1$) is often used to describe the so-called stretched relaxations of polymers and glasses. By contrast, our experimental results are satisfactorily fitted in the framework of the HN model with values $\alpha\approx 3/2 \div 1.7$ (i.e. greater than 1) and $\beta=1$. The HN relaxation model in Eq.(6) must satisfy general analytical conditions (causality) in the complex frequency plane $\omega$ [25]. In the case of $\alpha=3/2$ and $\beta=1$, the HN function and the distribution function (A1) have a branch point at $\omega=0, \tau_D=0$. The requirement of causality [25] will be satisfied if we make corresponding cuts in the line connecting two Riemann surfaces in the complex $\omega$ and $\tau_D$ planes. It is therefore difficult to interpret the function (A1) directly as due to a distribution of relaxation times in this case.

Instead, let us consider the HN relaxation function $F_{HN}(t)$ in the time-domain representation:

$$F_{HN}(t|\alpha=3/2,\beta=1) = \int_{-\infty}^{\infty} \frac{\exp(i\omega t)}{1+(i\omega\tau)^{3/2}} d\omega, \quad F_{HN}(t<0|\alpha=3/2,\beta=1) = 0 \quad (A2)$$

Expression (A2) was applied to the numerical calculation of the HN relaxation function shown in Fig.13. The function $F_{HN}(t>0|\alpha=3/2, \beta=1)$ can also be calculated by integration in the complex $\omega$-plane with cuts along the imaginary semi axis from $+0$ to $+i\infty$. We then have:

$$F_{HN}(t>0|\alpha=3/2, \beta=1) = -S + \frac{8\pi}{3\tau}\exp(-\frac{t}{2\tau})\cos(\frac{t}{\tau}\frac{\sqrt{3}}{2}-\frac{\pi}{3}),$$

$$S(t) = \frac{2}{t}\int_0^\infty \frac{(w\tau/t)^{3/2}}{(1+(w\tau/t)^3)}\exp(-w)dw,$$

$$S(t>>\tau) \approx \frac{2\tau^{3/2}}{t^{5/2}} \times \sum_{n=0}^{\infty}(-1)^n \Gamma(3n+5/2)!(\tau/t)^{3n} \quad (A3)$$

The pole term (A3) $\sim \exp(-\frac{t}{2\tau})\cos(\frac{t}{\tau}\frac{\sqrt{3}}{2}-\frac{\pi}{3})$ corresponds to the resonance at the frequency $\omega_p=\sqrt{(3/2/\tau)}$.

The parameters of this pole term (frequency, rate of decay) are in an agreement with those obtained from the Lorentzian fit in Fig.13.

APPENDIX B

Model of the tunneling transitions of the Nb off-center ion between (111) positions.

We consider tunneling transitions of one Nb ion between (111) type positions in the cubic unit cell. The Nb ion can occupy any one of the 8 sites at the corners of the unit cell with coordinates $x=\pm 1$, $y==\pm 1$, $z==\pm 1$. These are numbered as n=1-8. Their positions can be described by the (8x3) matrix S:

$S[n=1,x,y,z]=1,1,1; S[n=2,..]=-1,1,1; S[n=3,..]=-1,-1,1; S[n=4,..]=1,-1,1;$
$S[n=5,]=1,1,-1; S[n=6,..]=-1,1,-1; S[n=7,..]=-1,-1,-1; S[n=8,..]=1,-1,-1;$ (B1)

For simplicity we only take into account tunneling along cubic edges, i.e. between nearest (111) type positions. The tunneling Hamiltonian $H_T$ is written as an 8x8 matrix with corresponding eigenvectors $E(m)$ and orthogonal eigenfunctions $\psi(m)$ in the representation (B1):

$$H_T = -\Omega \begin{vmatrix} 0,1,0,1,1,0,0,0 \\ 1,0,1,0,0,1,0,0 \\ 0,1,0,1,0,0,1,0 \\ 1,0,1,0,0,0,0,1 \\ 1,0,1,0,0,0,0,1 \\ 1,0,0,0,0,1,0,1 \\ 0,1,0,0,1,0,1,0 \\ 0,0,1,0,0,1,0,1 \\ 0,0,0,1,1,0,1,0 \end{vmatrix} \quad (B2)$$

As an example, the ground state energy is $E=-3\Omega$ and corresponds to a completely symmetric wave function $\psi_S=[1,1,1,1,1,1,1,1]$. Wave functions corresponding to excited states are obtained by means of the commuting operators X, Y, Z:

$$X = \sum_{j=1}^{8} X^{(j)}, \quad Y = \sum_{j=1}^{8} Y^{(j)}, \quad Z = \sum_{j=1}^{8} Z^{(j)}, \quad XY = YX, \quad XZ = ZX, \quad XZ = ZX, \quad X^2 = Y^2 = Z^2 \sim 1 \quad (B3)$$

Triplet E=-$\Omega$, wave functions X$\psi_S$, Y$\psi_S$, Z$\psi_S$. Triplet E=+$\Omega$, wave functions XY$\psi_S$, YZ$\psi_S$, ZX$\psi_S$. Singlet E=+3 $\Omega$, wave function XYZ$\psi_S$. The dipole operators, X,Y, Z, excite transitions with energy ±2$\Omega$; quadruplet operators, XY,XZ, YZ, excite transitions with energy ±4$\Omega$ and *elastic* transitions within the triplet manifolds X$\psi_S$, Y$\psi_S$, Z$\psi_S$ and XY$\psi_S$, YZ$\psi_S$, ZX$\psi_S$; the octupole operator, XYZ, excites transitions ±2$\Omega$ ,±6$\Omega$.

The $Nb^{+5}$ ions are displaced from their high symmetry positions by $\boldsymbol{R_{Nb}}$ =0.145 A° in (111) directions. We express the values of the matrix elements in terms of the small dimensionless parameter $r_{Nb} \equiv R_{Nb}$/l.u. = 0.036 << 1; *X, Y, Z* ~ $r_{Nb}$ ; *XY, YZ, ZX*~ $r_{Nb}^2$ ; *XYZ* ~ $r_{Nb}^3$ .

The Hamiltonian terms describing the TO and TA phonon interaction with the tunneling ions will be different for short range (SR) and so-called long range (LR) forces which include polarization. For simplicity we do not take into account the modulation of the potential barrier by the TO and TA phonons.

1. TO, SR.

$\boldsymbol{R}^{(i)}$ designate the 8 polar vector operators, $\boldsymbol{\xi}$ the TO displacement and $\boldsymbol{n}$ is a unit vector directed along the TO momentum. The corresponding interaction *V(T0|SR)* can be written as follows:

$$V(TO|SR) \equiv V^{(1)}(TO|SR) + V^{(2)}(TO|SR),$$

$$V^{(1)}(TO|SR) = A(TO|dip)[\mathbf{n} \times \boldsymbol{\xi}] \sum_{i=1}^{i=8}[\mathbf{n} \times \mathbf{R}^{(i)}] \propto (r_{Nb}),$$

$$V^{(2)}(TO|SR) = A(TO|quad) \sum_{i,j=1}^{8} (R^{(i)}R^{(j)})_{\alpha\beta} (n\xi)_{\alpha\beta} \propto (r_{Nb})^2, \quad (B4)$$

$$(n\xi)_{\alpha,\beta} \equiv 1/2(n_\alpha \xi_\beta + n_\beta \xi_\alpha - 2/3 \delta_{\alpha\beta}(\mathbf{n}\boldsymbol{\xi})),$$

$$(R^{(i)}R^{(j)})_{\alpha,\beta} \equiv 1/2(R^{(i)}{}_\alpha R^{(j)}{}_\beta + R^{(i)}{}_\beta R^{(j)}{}_\alpha - 2/3 \delta_{\alpha\beta}(\mathbf{R}^{(i)}\mathbf{R}^{(j)}))$$

*A(TO|dip)* and *A(TO|quad)* are constants of interaction. The term $V^{(1)}$*(TO/SR)* describes dipole transitions with *ΔE=±2Ω*. $V^{(2)}$*(TO/SR)* excites transitions with energy *ΔE=±4Ω* and *elastic* transitions within the triplets *X$\psi_S$, Y$\psi_S$, Z$\psi_S$* and *XY$\psi_S$, YZ$\psi_S$, ZX$\psi_S$*. We should note that the $V^{(1)}$*(TO/SR)* dipolar interaction is much stronger than the quadrupolar one, $V^{(2)}$*(TO/SR)* .

2. TO, weak octupole transition.

$$V(TO|octo) \equiv A(TO|octo) \sum_{i,j,k=1}^{i,j,k=8} (\mathbf{n} \mathbf{R}^{(i)})(\mathbf{n} \mathbf{R}^{(j)})(\mathbf{R}^{(k)}\boldsymbol{\xi}) \propto (r_{Nb})^{+3} \quad (B5)$$

in which *A(TO|octo)* is the constant of the interaction. *V(TO|octo)* interaction could excite octuplet transitions with *ΔE=±6Ω* and dipole transitions with *ΔE=±2Ω* .

3. TA, Short Range (SR)

*V*(TA|SR) can be written as follows [30]:

$$V(TA|SR) = -\sum_{i=1}^{8} w^{(i)}{}_{\alpha\beta} \sigma_{\alpha,\beta} \quad (B6)$$

where $\sigma_{\alpha\beta}$ is the stress tensor and $w^{(i)}{}_{\alpha\beta}$ the elastic dipole *tensor* corresponding to the *i*-th (111) type cubic corner. The value of $w^{(i)}{}_{\alpha\beta}$ is defined by the geometry of the potential well. Let's suppose that $\lambda_L$ and $\lambda_P$ are the main values of the tensor $w^{(i)}{}_{\alpha\beta}$ in the coordinate system with z axis along a (111) direction. We then obtain in the coordinate system (100), (010), (001):

$$\sum_{i=1}^{i=8} w^{(i)} \sim \frac{2}{3}(\lambda_L - \lambda_P)(XY + YZ + ZX) \propto (r_{Nb})^2 \quad (B7)$$

Therefore the interaction of the *elastic* dipole with the stress generated by the TAW leads to quadrupolar transitions, $\Delta E = \pm 4\Omega$, and elastic transitions, $\Delta E = 0$, connecting states *inside* the triplets with energy $E = -\Omega$ and $E = +\Omega$.

4. TA, Long Range (LR).

Assuming that the electric field $E$ generated by the TA phonon and interacting with the electric dipole moment appears due to the displacement of the Nb$^{+5}$ ion from its high symmetry position, the interaction term, $V(TA|LR)$, can be written

$$V(TA|LR) = A_{TA} \sum_{i=1}^{8} (R^{(i)}P)_{\alpha\beta} \bar{u}_{\alpha,\beta} \sim, \quad \bar{u}_{\alpha,\beta} \equiv 1/2(u_\alpha q_\beta + u_\beta q_\alpha - 2/3\delta_{\alpha\beta}(\mathbf{uq}))$$

$$(R^{(i)}P) \equiv 1/2(R^{(i)}{}_\alpha P_\beta + R^{(i)}{}_\beta P_\alpha - 2/3\delta_{\alpha\beta}(R^{(i)}P)), \quad R^{(i)}{}_X = X^{(i)},... \quad (B8)$$

Here $A_{TA}$ is the constant of interaction and $P$ the polarization vector. Such polarization arises due the atomic displacements that accompany the formation of the PNDs. The physical meaning of (B8) can be explained as follows. Let's suppose that the polarization $P$ is directed along the z-axis. This perturbation creates a correction to the wave functions $\psi_S, X\psi_S$ such as

$$\delta\Psi_S = -\frac{P}{2\Omega} Z\Psi_S, \quad \delta X\Psi_S = --\frac{P}{2\Omega} ZX\Psi_S \quad (B9)$$

The elastic dipole-stress interaction (B7) and (B8) leads to the existence of the non-zero matrix element

$$\langle \Psi_S + \delta\Psi_S | V(TA|SR) | X\Psi_S + \delta X\Psi_S \rangle \sim \frac{P}{\Omega}(\lambda_L - \lambda_P)\sigma_{x,z}, \quad \Delta E = \pm 2\Omega \quad (B10)$$

Therefore the polarization "opens" transitions $\Delta E = \pm 2\Omega$ that are otherwise forbidden when $P=0$.

In the case of strong polarization, when $P \gg \Omega$, the wave function will be concentrated in the corners of the cubic unit cell in the plane z=-1, corresponding to tetragonal symmetry. The wave function components $S_P$ will then be:

$$S_P = [[1,1],[-1,1],[-1,-1],[1,-1]] \qquad (B11)$$

The tunneling Hamiltonian $H_{TP}$ and corresponding eigenvalues and eigenvectors in the representation (B11) can be written as follows:

$$H_{TP} = -\Omega \begin{bmatrix} 0,1,0,1 \\ 1,0,1,0 \\ 0,1,0,1 \\ 1,0,1,0 \end{bmatrix},$$

$$\Psi_{PS} = [1,1,1,1], \quad E_{PS} = -2\Omega,$$

$$\Psi_{PXmY} = (X-Y)/2 \, \Psi_{PS} = [0,-1,0,1], \quad E_{PXmY} = 0,$$

$$\Psi_{PXpY} = (X+Y)/2 \, \Psi_{PS} = [1,0,-1,0], \quad E_{PXpY} = 0, \qquad (B12)$$

$$\Psi_{PXY} = XY \Psi_{PS} = [1,-1,1,-1], \quad E_{PXY} = +2\Omega$$

The elastic dipole-stress interaction in (B6) and (B7) leads to the existence of non-zero matrix elements corresponding to the inelastic and elastic transitions, $\Delta E = \pm 2\Omega$ and $\Delta E = 0$, between states (B12) for the case when TAW is propagating along the z-axis. Transitions corresponding to $\Delta E = \pm 4\Omega$ could be excited for a TAW propagating and polarized in the X-Y plane. We should also note that the effect of the phonon polarization depends on the direction of the electric polarization $P$. As an example, for a strong polarization $P$ along the (111) axis, the wave function will be "locked" at the cubic corner (-1,-1,-1), and tunneling and the effect of the Nb off-centers on the damping of the TA phonon will be suppressed. In general, calculations corresponding to (B8)-(B12) are similar to the analysis of the features of a ferroelectric crystal placed in a symmetry-breaking field.